\documentclass[11pt]{article}
\usepackage[utf8]{inputenc}
\usepackage[T1]{fontenc}
\usepackage{lmodern}
\usepackage{microtype}
\usepackage{url}
\usepackage{hyperref}
\usepackage{amsmath,amssymb}
\usepackage{graphicx}
\usepackage{mathtools}
\usepackage{listings}
\usepackage{xcolor}

\lstdefinestyle{mystyle}{
    language=Matlab,
    backgroundcolor=\color{white},
    morekeywords={spmd},
    basicstyle=\ttfamily\small,
    keywordstyle=\color{blue},
    commentstyle=\color{green!60!black},
    stringstyle=\color{red},
    numbers=left,
    numberstyle=\tiny\color{gray},
    stepnumber=1,
    numbersep=8pt,
    frame=single,
    breaklines=true,
    breakatwhitespace=false,
    showstringspaces=false,
    tabsize=4
}

\title{Multi-GPU fast Fourier transforms in MATLAB \\ for large-scale phase-field crystal simulations}
\author{
Maik Punke\thanks{Institute of Scientific Computing, TU Dresden, 01062 Dresden, Germany}
\and
Marco Salvalaglio\footnotemark[1]  \thanks{Dresden Center for Computational Materials Science, TU Dresden, 01062 Dresden, Germany}
}
\date{\today}

\begin{document}
\maketitle

\section*{Summary}
We present a MATLAB-based framework for two- and three-dimensional fast Fourier transforms on multiple GPUs for large-scale numerical simulations using the pseudo-spectral Fourier method. The software implements two complementary multi-GPU strategies that overcome single-GPU memory limitations and accelerate spectral solvers. This approach is motivated by and applied to phase-field crystal (PFC) models, which are governed by tenth-order partial differential equations, require fine spatial resolution, and are typically formulated in periodic domains. Our resulting numerical framework achieves significant speedups, approximately sixfold for standard PFC simulations and up to sixtyfold for multiphysics extensions, compared to a purely CPU-based implementation running on hundreds of cores.

\section*{Statement of need}
Large-scale simulations based on the pseudo-spectral Fourier method \cite{boyd2001chebyshev} are often limited by the memory capacity and performance of single GPUs due to their reliance on repeated multidimensional fast Fourier transforms (FFTs). To address this bottleneck, we present two complementary multi-GPU FFT strategies in two and three dimensions.

The first strategy distributes a single high-dimensional FFT across multiple GPUs via domain decomposition and inter-GPU communication, enabling Fourier transforms for problem sizes that exceed the memory of individual GPUs. The second strategy computes multiple FFTs concurrently by assigning different physical fields to separate GPUs with synchronized communication.

In the following, phase-field crystal (PFC) models \cite{Elder2002,Elder2004,Emmerich2012} serve as a representative application that underscores the need for such multi-GPU capabilities. PFC models resolve crystalline order at atomic length scales while evolving on large (diffusive) time scales, enabling the simulation of elasticity, defects, grain boundaries, and microstructure evolution within a unified mesoscale framework. Capturing these phenomena typically requires relatively large periodic domains but fine spatial resolution, leading to substantial memory demands and limiting the reachable size of single-GPU execution. This is even more relevant for multiphysics extensions that involve additional variables, such as coupled density/composition, velocity, and/or temperature fields (see, e.g., \cite{skogvoll2022hydrodynamic,Punke_2022}).

The pseudo-spectral Fourier method is a natural and widely adopted numerical approach for PFC and related models, as it allows for efficient and accurate evaluation of high-order spatial derivatives and convolution operators that arise in their governing equations (see, e.g., \cite{cheng2008numeric,tegze2009advanced,cheng2019energy,Punke2023,punke2025hybrid,punke2026grain}). 
However, its performance is dominated by repeated multidimensional FFTs, which are required at every time step due to the nonlinear nature of the PDEs governing PFC dynamics, and quickly become the computational bottleneck at large scales. Efficient multi-GPU FFT strategies are therefore essential to fully exploit the accuracy and scalability of pseudo-spectral solvers.

As a representative use case, we thus exploit the developed strategies for multi-GPU execution in numerical simulations using a Fourier pseudo-spectral solver for the PFC model in two and three dimensions, where large spatial domains and coupled fields make single-GPU execution impractical. The algorithms are implemented in MATLAB, providing an accessible and extensible foundation for GPU-accelerated spectral simulations and introducing the first multi-GPU FFT implementation in MATLAB.

%These approaches are essential building blocks for pseudo-spectral solvers in large-scale and multiphysics applications. As a representative use case, we integrate both strategies into a Fourier pseudo-spectral solver for the phase-field crystal (PFC) model, where large spatial domains and coupled fields make single-GPU execution impractical. The algorithms are implemented in MATLAB, providing an accessible and extensible foundation for GPU-accelerated spectral simulations and introducing the first multi-GPU implementation in MATLAB.

\section*{Fourier pseudo-spectral method}
We briefly review the basics of the Fourier pseudo-spectral method. Although the actual implementation is carried out in two and three dimensions, we restrict the following discussion to a one-dimensional setting for the sake of simplicity. We consider a generic evolution equation for a scalar field $u\equiv u(x,t)$ of the form
\begin{equation}\label{eq:realdyn}
\partial_t u = \mathcal{L}u + \mathcal{N}(u), \quad x \in [0,2\pi), \ t \ge 0,
\end{equation}
where $\mathcal{L}$ is a linear differential operator, and $\mathcal{N}(u)$ is a nonlinear term given by a polynomial in $u$.

On the periodic domain $[0,2\pi)$, we approximate $u(x,t)$ by a truncated Fourier series
\begin{equation}
u(x,t) \approx u_N(x,t) = \sum_{k=-K}^{K} \widehat{u}_k  e^{ikx},
\end{equation}
with $N = 2K+1$ modes  $\widehat{u}_k\equiv \widehat{u}(k,t)$ and equispaced collocation points
\begin{equation}
x_j = \frac{2\pi j}{N}, \quad j = 0,\dots,N-1.
\end{equation}
In the Fourier pseudo-spectral method, a time-dependent ordinary differential equation for $\widehat{u}_k$ is obtained by applying the Fourier transform to the left and right-hand side of Eq.~\eqref{eq:realdyn}, reading
\begin{equation}\label{eq:fdyn}
\partial_t \widehat{u}_k = \widehat{\mathcal{L}}_k \widehat u_k + \widehat{[\mathcal{N}(u)]}_k \qquad \forall k=-K, \dots, K
\end{equation}
with $\widehat{\mathcal{L}}_k$ a polynomial in $k$ as $\widehat{[\partial^{n}_x u]}_k = (ik)^n \widehat{u}_k$, $\forall n\in \mathbb{N}$ while the nonlinear polynomial term is evaluated point-wise in physical space at every time iteration and then transformed to Fourier space. For a given $u(x,t)$ and time step $\Delta t$,  $u(x,t+\Delta t)$ is computed by first determining $\widehat{u}_k(t)$ and $\widehat{[\mathcal{N}(u)]}_k(t)$ via FFTs, solving the system of ODEs~\eqref{eq:fdyn} using a suitable time-integration scheme, and then computing the inverse FFT from the resulting $\widehat{u}_k(t+\Delta t)$. The concept can be readily extended to higher spatial dimensions. More details can be found in Ref.~\cite{boyd2001chebyshev}.

\section*{Single FFT on Multiple GPUs}

Distributed multi-GPU FFTs based on domain decomposition and inter-GPU communication are well established in high-performance computing; see, e.g.,~\cite{pekurovsky2012p3dfft,ayala2020heffte,verma2023scalable,cuFFTmp}. These approaches enable scalable multi-dimensional FFTs on large GPU systems.
In contrast, FFT-based pseudo-spectral solvers for the PFC equation have so far been limited to single-GPU implementations~\cite{hallberg2025pypfc} or CPU-based parallelization~\cite{pinomaa2024openpfc,skogvoll2024comfit}. To our knowledge, a dedicated multi-GPU FFT implementation tailored to large-scale three-dimensional PFC simulations has not been reported. Furthermore, a general multi-GPU FFT implementation is not yet available in MATLAB, independent of any specific application such as PFC simulations.

\begin{figure*}[t]
\centering
\includegraphics[width=\linewidth]{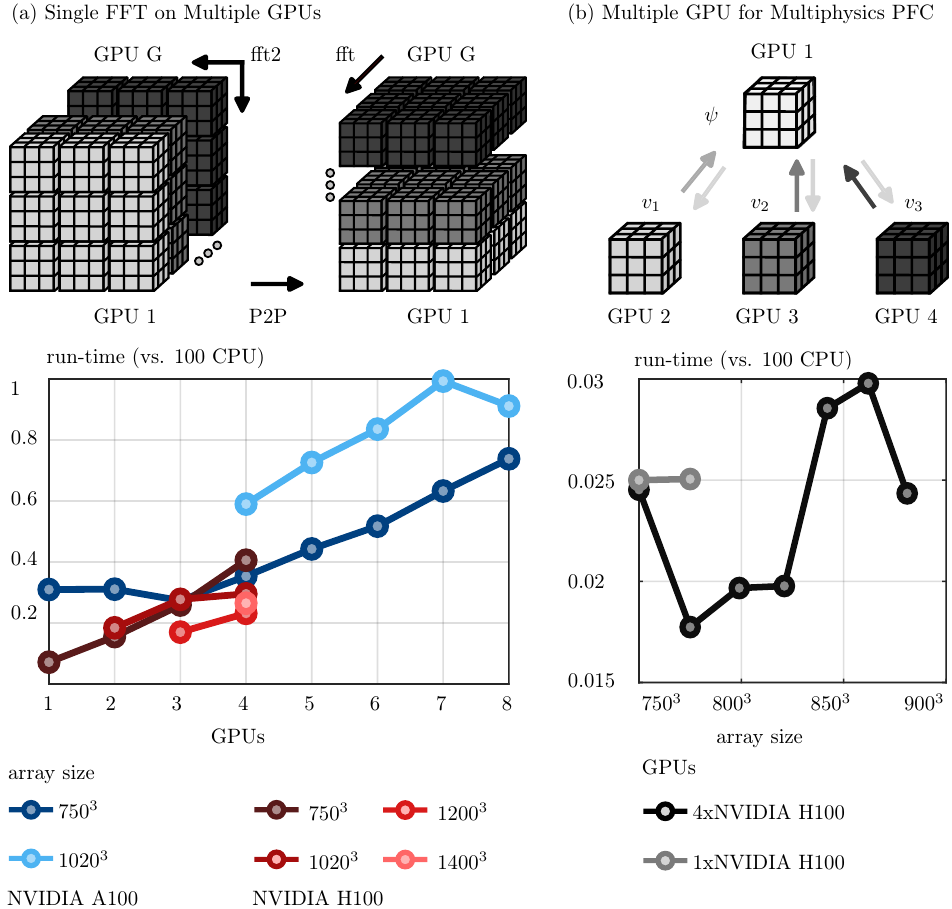}
\caption{
(a) Schematic of the multi-GPU FFT algorithm based on slab decomposition for a three-dimensional array. The data are decomposed along the $z$-direction, followed by local two-dimensional FFTs, peer-to-peer communication, and a final one-dimensional FFT (upper panel). Relative runtimes for $1000$ time steps are shown, normalized by CPU execution time. Speedups of up to a factor of six are observed, with optimal performance on a single GPU for $750^3$ and multi-GPU execution required for larger domains due to memory constraints. An array of size $1400^3$ fits only on four H100 GPUs and reaches approximately 17 \% of the CPU runtime (lower panel).
(b) Decomposition of a pseudo-spectral multiphysics PFC solver (e.g., hydrodynamic PFC) across four GPUs (upper panel). Relative runtimes of the hydrodynamic PFC solver are shown as a function of problem size, demonstrating that multi-GPU execution enables simulations up to a problem size of $900^3$, which is infeasible on a single GPU. Compared to a CPU implementation, speedups of up to 60$\times$ are achieved (lower panel).
}
	\label{fig:multiGPU}
\end{figure*}

Following the strategy of \cite{pekurovsky2012p3dfft, ayala2020heffte}, we decompose a three-dimensional FFT of size $N_x \times N_y \times N_z$ across $G$ GPUs by slab decomposition along the $z$-direction, cf. Fig.~\ref{fig:multiGPU}(a). Each GPU first performs local two-dimensional FFTs, then performs peer-to-peer (P2P) communication to redistribute the data, and finally performs a final one-dimensional FFT along the remaining dimension. After completion, each GPU holds a portion of the transformed Fourier-space array.

We apply this strategy to the PFC equation, which describes the evolution of a periodic density field $\psi\equiv \psi(\mathbf{x},t)$~\cite{Elder2002,Elder2004}. The model builds on a free-energy functional, which, for example, can be expressed for face-centered cubic (FCC) crystal symmetry as
\begin{equation}
\label{eq:SHenergy}
F[\psi]
= \int_\Omega \left[
\frac{\psi}{2}\left(\varepsilon + \mathcal{L}\right)\psi
+ \frac{\psi^4}{4}
\right]  \mathrm{d}\mathbf r,
\end{equation}
with undercooling parameter $\varepsilon$ and  $\mathcal{L} = (1+\nabla^2)^2(4/3+\nabla^2)^2$ an operator describing spatial correlations within the  domain $\Omega$. The classical evolution equation reads
\begin{equation}
\label{eq:evolution}
\partial_t \psi
=\nabla^2 \dfrac{\delta F[\psi]}{\delta \psi}=(\varepsilon + \mathcal{L}) \nabla^2 \psi
+ \nabla^2 \psi^3
\end{equation}
which retains the form of Eq.~\eqref{eq:realdyn}. The equation is solved using a Fourier pseudo-spectral method with semi-implicit time integration~\cite{tegze2009advanced}. We note that alternative time-integration approaches can be considered~\cite{Punke2023}.

We benchmark the multi-GPU solver for problem sizes ranging from $750^3$ to $1400^3$, achieving up to a sixfold speedup compared to a purely CPU-based implementation; see Fig.~\ref{fig:multiGPU}(a). The benchmarks are performed on three systems from the HPC clusters provided by the NHR Center at TU Dresden. GPU computations are performed with four NVIDIA H100 SXM5 GPUs (94~GiB HBM2e each, HPC cluster Capella) and with eight NVIDIA A100 SXM4 GPUs (40~GiB HBM2 each, HPC cluster Alpha Centauri). CPU reference runs are performed on an Intel Xeon Platinum~8470 (100 cores, 2.00~GHz, HPC cluster Barnard). In Listing~\ref{lst:matlab1}, the MATLAB implementation is illustrated.

\begin{figure*}[t]
\centering
\includegraphics[width=\linewidth]{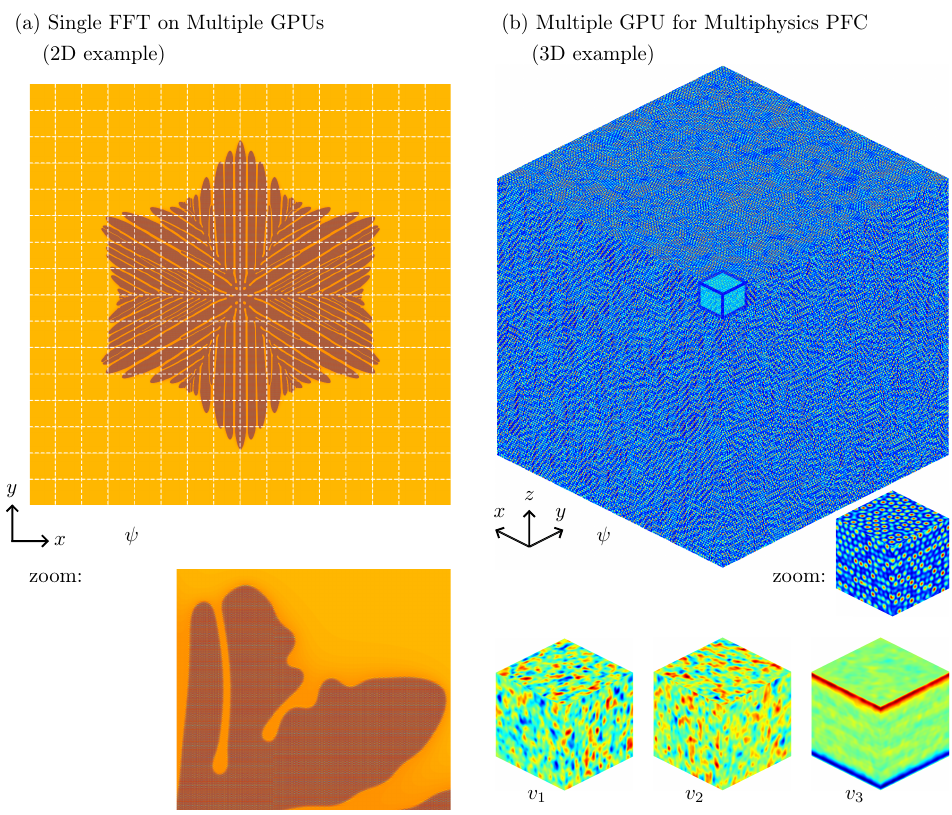}
\caption{
Representative large-scale PFC benchmark problems for multi-GPU FFT algorithms:
(a) Dendritic solidification (underlying triangular crystal symmetry) using the multi-GPU single-FFT implementation (2D example; computational domain of size $5\cdot 10^4\times 5\cdot 10^4$, corresponding to $2.5 \,\mu\mathrm{m}\times 2.5 \,\,\mu\mathrm{m}$ when assuming a lattice constant of $4\AA$ for aluminum). The density field $\psi$ along with a close-up is shown. White lines delineate patches, as the array exceeds single-plot size limits.
(b) Polycrystalline coarsening of an FCC crystal structure using the multi-GPU hydrodynamic PFC solver. Visualized are the density field $\psi$ (including a magnified view) and the velocity components $\mathbf{v}_1$, $\mathbf{v}_2$, $\mathbf{v}_3$. A grid of $1400\times 1400\times 1400$ is used which corresponds to a box size of $40\,\mathrm{nm}\times 40\,\mathrm{nm}\times 40\,\mathrm{nm}$. The material and model parameters are documented in the repository accompanying this work.
}
	\label{fig:multiGPU_examples}
\end{figure*}

Figure \ref{fig:multiGPU_examples}(a) illustrates dendritic solidification within the PFC framework, presented here as a representative two-dimensional benchmark example.

\section*{Multiple GPU usage for Multiphysics PFC}
The PFC framework readily supports multiphysics extensions. As an example, we consider the hydrodynamic phase-field crystal (hydrodynamic PFC) model~\cite{skogvoll2022hydrodynamic,qiu2024grain} in three spatial dimensions, which augments the density field $\psi$ with a mesoscopic velocity field $\mathbf{v} \equiv (\mathbf{v}_1(\mathbf{x},t),\mathbf{v}_2(\mathbf{x},t),\mathbf{v}_3(\mathbf{x},t))$:
\begin{equation}
\label{eq:hpfc}
\begin{aligned}
\partial_t \psi &= \nabla^2\left(\frac{\delta F[\psi]}{\delta \psi}\right) -\mathbf{v}  \cdot \nabla \psi, \\
\rho \partial_t \mathbf{v} &= \Gamma \nabla^2 \mathbf{v}
- \Big\langle \psi \nabla\frac{\delta F[\psi]}{\delta \psi} \Big\rangle ,
\end{aligned}
\end{equation}
with $\langle\ \cdot \ \rangle$ a local averaging obtained through a convolution with a Gaussian kernel 
\begin{equation}\label{eq:coarsegraining}
\langle \ \cdot\ \rangle(\mathbf{r})=\int_\Omega  \frac{ (\ \cdot\ )\left(\mathbf{r}^{\prime}\right)}{\left(2 \pi a_0^2\right)^{3 / 2}} \exp \left(-\frac{\left(\mathbf{r}-\mathbf{r}^{\prime}\right)^2}{2 a_0^2}\right) d \mathbf{r}^{\prime}
\end{equation}
and $a_0$ the lattice spacing. Note that this operation just translates to a multiplication in the Fourier spectral method owing to the properties of the Fourier transform.

We distribute the fields across four GPUs, assigning one field per device and performing synchronized inter-GPU communication after each time step. This strategy enables handling problems larger than single-GPU memory limits. Benchmark results using four NVIDIA H100 GPUs demonstrate substantial runtime reductions compared to CPU execution on the previously described HPC clusters Capella and Barnard. In particular, for large computational domains where single-GPU simulations become infeasible, significant performance gains are observed; see Fig.~\ref{fig:multiGPU}(b). Listing~\ref{lst:matlab2} illustrates the MATLAB implementation. Figure~\ref{fig:multiGPU_examples}(b) depicts polycrystalline coarsening within the hydrodynamic PFC framework, presented here as a representative three-dimensional benchmark case.

The adopted strategy of distributing different fields across multiple GPUs naturally extends to coarse-grained formulations based on complex amplitudes of the principal Fourier modes \cite{salvalaglio2022coarse}. These models are well-suited to pseudo-spectral methods and typically require the simultaneous evolution of tens of coupled complex-valued fields, a setting for which we expect the present parallelization strategy to be particularly effective.

\newpage

\section*{Implementation Snippet}

\begin{lstlisting}[caption={Single FFT on Multiple GPUs}, label={lst:matlab1}]
%%initialize and decompose into slabs
% psi ...initial density field
% psiF...Fourier transformed initial density field
% lap ...discretized laplacian
% lin ...discretized linear operator (epsilon+L^2)nabla^2

spmd %parallel GPU session (G GPUs), on each GPU perform
    for  n=1:n_timeSteps %time iteration
        %%forward fftn
        psi = fft2(psi.^3); %fft2 along first two dimensions
        for i=1:G %P2P communication 
            psi_chunk = psi(localStart:localStart,:,:); %decompose each slab in a set of local chunks
            psi_chunk= spmdCat(psi_chunk,3,i); %stack the chunks together (along third dimension)
        end
        psi = psi_chunk;
        psi = fft(psi,[],3);  %fft along third dimension
        
        %%semi-implicit time step update
        psiF = (psiF+dt*lap.*psi)./(1-dt*lin); 

        %%backward ifftn
        psi = psiF;
        psi  = ifft(psi,[],3); %ifft along third dimension
        for i=1:G %P2P communication  
            psi_chunk = psi(:,:,localStart:localStart);%decompose each slab in a set of local chunks
            psi_chunk= spmdCat(psi_chunk,1,i); %stack the chunks together (along first dimension)
        end
        psi = psi_chunk;
        psi  = ifft2(psi); %ifft2 along first two dimensions
    end
end

%% stack solution 
psi = cat(3, gather(psi{:}));
\end{lstlisting}

\begin{lstlisting}[caption={Multiple GPU usage for Multiphysics PFC}, label={lst:matlab2}]
%%initialize on GPU1
% psi ...initial density field
% psiF...Fourier transformed initial density field
% v1  ...initial velocity field (first component)
% v2  ...initial velocity field (secod component)
% v3  ...initial velocity field (third component)
% lin ...discretized linear operator (epsilon+L^2)nabla^2
% k   ...discretized Fourier vector (first dimension)
% l   ...discretized Fourier vector (second dimension)
% m   ...discretized Fourier vector (third dimension)

%%initialize on GPU2...GPU4
% lap ...discretized laplacian
% k   ...discretized Fourier vector (first dimension)
% op  ...discretized linear operator (epsilon+L^2)
% cg  ... discretized convolution kernel
% additionally initialize on GPU2
% v1F ...Fourier transformed initial v1
% additionally initialize on GPU3
% v2F ...Fourier transformed initial v2
% additionally initialize on GPU4
% v3F ...Fourier transformed initial v3

spmd %parallel GPU session (4 GPUs)
    for  n=1:n_timeSteps %time iteration
        if spmdIndex==1 %GPU1 holding psi
            %%semi-implicit time step update of psi
            psiF = (psiF+dt*(lap.*fftn(psi.^3)-fftn(v1.*ifftn(k.*psiF)+v2.*ifftn(l.*psiF)+v3IF.*ifftn(m.*psiF))))./(1-dt*lin);
            psi = ifftn(psiF);
            %%send psi to GPU2...GPU4
            spmdSend(psi,[2:4],2); 
            %%receive v components from GPU2...GPU4
            v1 = spmdReceive(2,4);
            v2 = spmdReceive(3,5);
            v3 = spmdReceive(4,6);
            
        elseif spmdIndex==2 %GPU2 holding v1
            %receive psi from GPU1
            psi = spmdReceive(1,2);
            %%fully-implicit time step update of v1
            v1F = (v1F-dt/rho0 .*cg.*fftn(psi.*ifftn(k.*(fftn(psi.^3) + op.*fftn(psi)))))
                ./(1-deltatEuler/rho0.*.GammaS.*lap));
            v1 = ifftn(v1F);
            %send v1 to GPU1
            spmdSend(v1,1,4);
            
        elseif spmdIndex==3 %GPU3 holding v2
            %receive psi from GPU1
            psi = spmdReceive(1,2);
            %%fully-implicit time step update of v2
            v2F = (v2F-dt/rho0 .*cg.*fftn(psi.*ifftn(l.*(fftn(psi.^3) + op.*fftn(psi)))))
                ./(1-deltatEuler/rho0.*.GammaS.*lap));
            v2 = ifftn(v2F);
            %send v2 to GPU1
            spmdSend(v2,1,5);
            
        elseif spmdIndex==4 %GPU4 holding v3
            %receive psi from GPU1
            psi = spmdReceive(1,2);
            %%fully-implicit time step update of v3
            v3F = (v3F-dt/rho0 .*cg.*fftn(psi.*ifftn(m.*(fftn(psi.^3) + op.*fftn(psi)))))
                ./(1-deltatEuler/rho0.*.GammaS.*lap));
            v3 = ifftn(v3F);
            %send v3 to GPU1
            spmdSend(v3,1,6);
        end    
    end
end
\end{lstlisting}

\section*{Availability}
\textbf{Repository:} \url{https://github.com/mpunke/MATLABmultiGPUFFT/} \\
\textbf{License:} MIT License \\
\textbf{Zenodo DOI:} \url{https://doi.org/10.5281/zenodo.18670913}

\section*{Acknowledgements}
We acknowledge support from the Deutsche Forschungsgemeinschaft (DFG, German Research Foundation), project numbers 447241406, 493401063. We also gratefully acknowledge the computing on the high-performance computer at the NHR Center of TU Dresden. This center is jointly supported by the Federal Ministry of Education and Research and the state governments participating in the NHR (\href{www.nhr-verein.de/unsere-partner}{www.nhr-verein.de/unsere-partner}).

\bibliographystyle{refs_style} 
\bibliography{refs}

@article{punke2025hybrid,
  title={Hybrid-PFC: Coupling the phase-field crystal model and its amplitude-equation formulation},
  author={Punke, Maik and Salvalaglio, Marco},
  journal={Computer Methods in Applied Mechanics and Engineering},
  volume={436},
  pages={117719},
  year={2025},
  publisher={Elsevier},
  doi ={10.1016/j.cma.2024.117719}
}

@article{skogvoll2022hydrodynamic,
  title={Hydrodynamic phase field crystal approach to interfaces, dislocations, and multi-grain networks},
  author={Skogvoll, Vidar and Salvalaglio, Marco and Angheluta, Luiza},
  journal={Modelling and Simulation in Materials Science and Engineering},
  volume={30},
  number={8},
  pages={084002},
  year={2022},
  publisher={IOP Publishing},
  doi = {10.1088/1361-651X/ac9493}
}

@article{skogvoll2024comfit,
  title={ComFiT: a Python library for computational field theory with topological defects},
  author={Skogvoll, Vidar and R{\o}nning, Jonas},
  journal={Journal of Open Source Software},
  volume={9},
  number={98},
  pages={6599},
  year={2024},
  doi = {10.21105/joss.06599}
}

@article{cheng2019energy,
  title={An Energy Stable BDF2 Fourier Pseudo-Spectral Numerical Scheme for the Square Phase Field Crystal Equation},
  author={Cheng, Kelong and Wang, Cheng and Wise, Steven M},
  journal={Communications in Computational Physics},
  volume={26},
  number={5},
  pages={1335--1364},
  year={2019},
  doi={10.4208/cicp.2019.js60.10}
}

@book{boyd2001chebyshev,
  title={Chebyshev and Fourier spectral methods},
  author={Boyd, John P},
  year={2001},
  publisher={Courier Corporation}
}

@article{cheng2008numeric,
title = {An efficient algorithm for solving the phase field crystal model},
journal = {Journal of Computational Physics},
volume = {227},
number = {12},
pages = {6241-6248},
year = {2008},
doi = {https://doi.org/10.1016/j.jcp.2008.03.012},
author = {Mowei Cheng and James A. Warren}
}

@article{qiu2024grain,
  title={Grain boundaries are Brownian ratchets},
  author={Qiu, Caihao and Punke, Maik and Tian, Yuan and Han, Ying and Wang, Siqi and Su, Yishi and Salvalaglio, Marco and Pan, Xiaoqing and Srolovitz, David J and Han, Jian},
  journal={Science},
  volume={385},
  number={6712},
  pages={980--985},
  year={2024},
  publisher={American Association for the Advancement of Science},
  doi = {10.1126/science.adp1516}
}

@article{salvalaglio2022coarse,
  author={Salvalaglio, Marco and Elder, Ken R},
  journal={Model. Simul. Mater. Sci. Eng.},
doi = {10.1088/1361-651x/ac681e},
	year = 2022,
	month = {may},
	publisher = {{IOP} Publishing},
	volume = {30},
	number = {5},
	pages = {053001},
	title = {Coarse-grained modeling of crystals by the amplitude expansion of the phase-field crystal model: an overview}
}

@inproceedings{ayala2020heffte,
  title={heffte: Highly efficient fft for exascale},
  author={Ayala, Alan and Tomov, Stanimire and Haidar, Azzam and Dongarra, Jack},
  booktitle={International Conference on Computational Science},
  pages={262--275},
  year={2020},
  organization={Springer},
  doi = {10.1007/978-3-030-50371-0_19}
}

@article{Punke2023,
author = {Punke, Maik and Wise, Steven M. and Voigt, Axel and Salvalaglio, Marco},
title = {Improved time integration for phase-field crystal models of solidification},
journal = {PAMM},
volume = {23},
number = {1},
pages = {e202200112},
doi = {https://doi.org/10.1002/pamm.202200112},
year = {2023}
}

@article{Elder2004,
author = {Elder, K. R. and Grant, Martin},
journal = {Phys. Rev. E},
number = {5},
pages = {051605},
volume = {70},
doi = {10.1103/PhysRevE.70.051605},
year = {2004}
}

@article{Punke_2022,
	doi = {10.1088/1361-651x/ac8abd},
	year = 2022,
	month = {sep},
	publisher = {{IOP} Publishing},
	volume = {30},
	number = {7},
	pages = {074004},
	author = {Maik Punke and Steven M Wise and Axel Voigt and Marco Salvalaglio},
	title = {Explicit temperature coupling in phase-field crystal models of solidification},
	journal = {Modelling and Simulation in Materials Science and Engineering}
}

@article{Emmerich2012,
author = {Emmerich, Heike and L{\"{o}}wen, Hartmut and Wittkowski, Raphael and Gruhn, Thomas and T{\'{o}}th, Gyula I and Tegze, Gy{\"{o}}rgy and Gr{\'{a}}n{\'{a}}sy, L{\'{a}}szl{\'{o}}},
doi = {10.1080/00018732.2012.737555},
journal = {Adv. Phys.},
number = {6},
pages = {665--743},
title = {{Phase-field-crystal models for condensed matter dynamics on atomic length and diffusive time scales: an overview}},
volume = {61},
year = {2012}
}

@article{pinomaa2024openpfc,
  title={OpenPFC: an open-source framework for high performance 3D phase field crystal simulations},
  author={Pinomaa, Tatu and Aho, Jukka and Suviranta, Jaarli and Jreidini, Paul and Provatas, Nikolas and Laukkanen, Anssi},
  journal={Modelling and Simulation in Materials Science and Engineering},
  volume={32},
  number={4},
  pages={045002},
  year={2024},
  publisher={IOP Publishing},
  doi = {10.1088/1361-651X/ad269e}
}

@article{hallberg2025pypfc,
  title={pyPFC: an open-source Python package for phase field crystal simulations},
  author={Hallberg, H{\aa}kan and Blixt, Kevin H},
  journal={Modelling and Simulation in Materials Science and Engineering},
  volume={34},
  number={1},
  pages={015004},
  year={2025},
  publisher={IOP Publishing},
  year={2026},
  doi = {10.1088/1361-651X/ae2599}
}

@article{pekurovsky2012p3dfft,
  title={P3DFFT: A framework for parallel computations of Fourier transforms in three dimensions},
  author={Pekurovsky, Dmitry},
  journal={SIAM Journal on Scientific Computing},
  volume={34},
  number={4},
  pages={C192--C209},
  year={2012},
  publisher={SIAM},
  doi     = {10.1137/11082748X}
}

@article{Elder2002,
author = {Elder, K R and Katakowski, Mark and Haataja, Mikko and Grant, Martin},
journal = {Phys. Rev. Lett.},
pages = {245701},
title = {{Modeling Elasticity in Crystal Growth}},
volume = {88},
doi = {https://doi.org/10.1103/PhysRevLett.88.245701},
year = {2002}
}

@article{verma2023scalable,
  title={Scalable multi-node fast fourier transform on GPUs},
  author={Verma, Manthan and Chatterjee, Soumyadeep and Garg, Gaurav and Sharma, Bharatkumar and Arya, Nishant and Kumar, Sashi and Saxena, Anish and Verma, Mahendra K},
  journal={SN Computer Science},
  volume={4},
  number={5},
  pages={625},
  year={2023},
  publisher={Springer},
  doi     = {10.1007/s42979-023-02109-0}
}

@misc{cufftmp,
  author       = {{NVIDIA Corporation}},
  title        = {cuFFTMp: Distributed FFTs on Multi-GPU and Multi-Node Systems},
  year         = {2023},
  howpublished = {\url{https://developer.nvidia.com/cufft}}
}

@article{punke2026grain,
  title={Grain-Growth Stagnation from Vacancy-Diffusion-Limited Disconnection Climb},
  author={Punke, Maik and Milor, Abel HG and Salvalaglio, Marco},
  journal={arXiv preprint arXiv:2601.13965},
  year={2026},
  doi = {10.48550/arXiv.2601.13965}
}

@article{tegze2009advanced,
  title={Advanced operator splitting-based semi-implicit spectral method to solve the binary phase-field crystal equations with variable coefficients},
  author={Tegze, Gy{\"o}rgy and Bansel, Gurvinder and T{\'o}th, Gyula I and Pusztai, Tam{\'a}s and Fan, Zhongyun and Gr{\'a}n{\'a}sy, L{\'a}szl{\'o}},
  journal={Journal of Computational Physics},
  volume={228},
  number={5},
  pages={1612--1623},
  year={2009},
  publisher={Elsevier},
  doi     = {10.1016/j.jcp.2008.11.011}
}

\end{document}